\documentclass[12pt]{article}

\usepackage{amsmath,amsfonts}
\usepackage{amsthm}
\usepackage[mathscr]{eucal}

\newcommand{\afs}{\mathbf{A}}
\newcommand{\bra}[1]{\pmb{\langle}#1\pmb{|}}

\newcommand{\cdt}{,\ldots,}
\newcommand{\dff}{\sc}
\newcommand{\hhh}{\mathcal{H}}
\newcommand{\hhhh}{\mathcal{H}}
\newcommand{\ket}[1]{\pmb{|}#1\pmb{\rangle}}
\newcommand{\kkk}{\mathcal{K}}
\newcommand{\lth}{N}

\newcommand{\ptt}{\sigma}
\newcommand{\pttr}{\Pi}
\newcommand{\pttt}{\Sigma}
\newcommand{\rrh}{\mathbf{\rho}}
\newcommand{\sss}{\mathscr{S}}
\newcommand{\uu}{\mathbf{U}}
\newcommand{\vvv}{V}
\newcommand{\yyyy}{\mathbf{S}}
\newcommand{\yyyyy}{\mathbf{A}}

\DeclareMathOperator{\relloc}{\mathbf{U}}

\newcommand{\ketbra}[2]{\ket{#1}\bra{#2}}
 
\newcommand{\ensm}{{\mathbf{E}}}
\title{Combinatorial topology of multipartite entangled states}

\author{
Rom\`an R. Zapatrin\\
\small\itshape 
Quantum Information Group, ISI, 
Villa Gualino,\\ 
\small\itshape 
Viale Settimio Severo 65,
10133, Torino, Italy;\\ 
\small\rm
e-mail: zapatrin@isiosf.isi.it
}

\begin{document}

\maketitle

\begin{abstract}
With any state of a multipartite quantum system its separability 
polytope is associated. This is an algebro-topological object 
(non-trivial only for mixed states) which captures the localisation 
of entanglement of the state. Particular examples of separability 
polytopes for 3-partite systems are explicitly provided. It turns  
out that this characterisation of entanglement is associated with 
simulation of arbitrary unitary operations by 1- and 2-qubit gates. 
A topological description of how entanglement changes  in course of 
such simulation is provided. 
\end{abstract}


\section*{Introduction}

Entanglement in multipartite quantum system is now treated as a key 
resource in quantum information processing. That is why multiple 
efforts are drawn to quantification of entanglement for quantum 
states. 

For bipartite quantum systems all entanglement measures are 
essentially of numerical nature as a single real positive number is 
enough to quantify the degree of entanglement. In the case of 
multipartite systems the situation differs drastically. It was 
observed that even pure states of a 3-particle system can be 
entangled in different ways, which can not be interconverted by 
local unitary transformations. 

In this paper, starting from the classification of mixed bipartite 
states \cite{durcirac} I introduce a non-numerical, topological 
characterisation of mixed $\lth$-partite quantum states. It is 
known that, in contrast with the classical case, a quantum state 
can be represented by different inequivalent ensembles of pure 
states. I consider equivalence classes of ensembles with respect to 
their localisation. With each state a polytope (a geometrical 
object in an affine space) is associated such that every maximal 
simplex of the polytope represents a class of equivalently 
localised ensembles.  

To make the paper complete, all the necessary definitions are 
provided. 

\section{Composite systems and partitions}

Let $\yyyy$ be a $\lth$-partite physical system. We may dwell on a 
less detailed description of multipartite nature of $\yyyy$, 
namely, grouping subsystems in their most detailed setting and 
considering these groups as  subsystems. Each such passage to a 
less detailed description of the composite structure of $\yyyy$ we 
associate with appropriate \underline{partition} of the total set 
$\yyyyy$ of subsystems of $\yyyy$. 

Denote the collection of all partitions of $\yyyyy$ by 
$\Pi(\yyyyy)$ or $\Pi(\lth)$ where $\lth$ is the  number of the 
subsystems. The first observation is that partitions are ordered. 
A partition $\pttt$ is called {\dff  finer} than a partition 
$\pttt'$ (and $\pttt'$ is, respectively, {\dff coarser} than 
$\pttt$ whenever any element of $\pttt'$ is a union of 
elements of the  partition $\pttt$. Denote it 

\begin{equation}\label{e01}
\pttt'\preceq\pttt
\end{equation}

\medskip 

The order relation $\preceq$ on partitions is partial, that is, two 
partitions may be  incomparable. To elucidate it, consider an 
example. 

\begin{figure}[h!]\label{fig1}
\begin{center}
\unitlength1mm
$\pttt$
\;=\;
\begin{picture}(20,20)(0,9)
\put(0,0){\line(1,0){20}}
\put(0,20){\line(1,0){20}}
\put(0,0){\line(0,1){20}}
\put(20,0){\line(0,1){20}}
\put(4,14){\mbox{1}}
\put(14,14){\mbox{2}}
\put(4,4){\mbox{3}}
\put(14,4){\mbox{4}}
\put(0,10){\line(1,0){20}}
\put(10,0){\line(0,1){10}}
\end{picture}
\vspace{10mm}
\;;\quad
$\pttt'$
\;=\;
\begin{picture}(20,20)(0,9)
\put(0,0){\line(1,0){20}}
\put(0,20){\line(1,0){20}}
\put(0,0){\line(0,1){20}}
\put(20,0){\line(0,1){20}}
\put(4,14){\mbox{1}}
\put(14,14){\mbox{2}}
\put(4,4){\mbox{3}}
\put(14,4){\mbox{4}}
\put(0,10){\line(1,0){20}}
\end{picture}
\;;\quad
$\pttt''$
\;=\;
\begin{picture}(20,20)(0,9)
\put(0,0){\line(1,0){20}}
\put(0,20){\line(1,0){20}}
\put(0,0){\line(0,1){20}}
\put(20,0){\line(0,1){20}}
\put(4,14){\mbox{1}}
\put(14,14){\mbox{2}}
\put(4,4){\mbox{3}}
\put(14,4){\mbox{4}}
\put(10,0){\line(0,1){20}}
\end{picture}
\end{center}
\caption{An example of different partitions of a 4-partite 
system.} 
\end{figure}
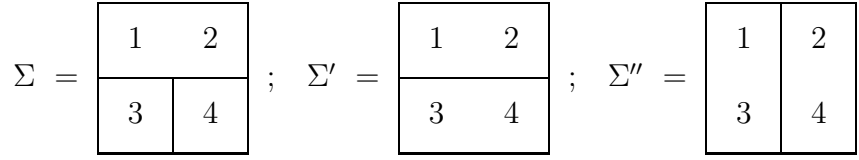

\paragraph{Example.} Let $\yyyy$ be a 4-partite system, enumerate 
its components by numbers $\yyyyy=\{1,2,3,4\}$. Consider the 
following partitions $\pttt=\{12,3,4\}, \pttt'=\{12,34\}, 
\pttt''=\{13,24\}$, see figure \ref{fig1}.

\medskip 

\noindent For instance, we treat $(\yyyy,\pttt)$ as a 
tripartite system for which 1 and 2 together are treated as one 
subsystem while $(\yyyy,\pttt')$ is bipartite. Among these  
partitions $\pttt'\preceq\pttt$, while $\pttt''$ is not comparable  
with any of $\pttt, \pttt'$. 

\medskip 

To summarise, we observe that besides the initial, finest 
decomposition of the state space $\hhhh$ of the composite system 
$\yyyy$ we may consider coarser decompositions each of which 
associated with certain partition of the set $\yyyyy$ of subsystems 
of $\yyyy$. Let us denote it as  follows. The initial decomposition 
of $\hhhh$ is 

\[
\hhhh
=
\otimes_{i\in\yyyyy}
\hhh_i
\]

\noindent and each partition $\pttt=\{\ptt_1,\ldots\ptt_M\}$ (where 
$\ptt_1\cup\cdots\cup\ptt_M=\yyyyy$) gives rise to the following 
decomposition of the overall state space $\hhhh$ as:

\begin{equation}\label{e02}
\hhhh
=
\hhh_{\ptt_1}
\otimes\cdots\otimes
\hhh_{\ptt_M}
\end{equation}

\subsection{Product, separable and entangled states.}

Given a state of a composite $\lth$-partite system $\yyyy$, denote 
its density matrix by $\rrh$. A density matrix $\rrh$ is called 
{\dff product} if it can be represented as a tensor product of 
density matrices of subsystems 

\begin{equation}\label{e03}
\rrh
=
\rrh_1
\otimes\ldots\otimes
\rrh_\lth
\end{equation}

\noindent A state $\rrh$ is {\dff separable} if its density matrix 
is a convex combination of product ones. 

\begin{equation}\label{e03s}
\rrh
=
\sum
p_\alpha
\rrh_1^\alpha
\otimes\ldots\otimes
\rrh_\lth^\alpha
\end{equation}

\noindent with $p_\alpha\ge0$ and $\sum{}p_\alpha=1$. 

\medskip 

The states which are not separable are called {\dff entangled}. 

\medskip 

\subsection{The relativity of multipartite entanglement} 

Now let us weaken the condition for states of a composite system 
$\yyyy$ to be product and separable. Namely, instead of 
requiring \eqref{e03} for the tensor product 

\[
\hhhh
=
\hhh_1
\otimes\cdots\otimes
\hhh_\lth
\] 

\noindent we make this condition \underline{relative} with respect 
to a partition $\pttt$ of the set $\yyyyy$ of subsystems of 
$\yyyy$, that is, with respect to the decomposition \eqref{e02}. 

\medskip 

Given a partition $\pttt=\{\ptt_1,\ldots,\ptt_M\}$ of 
$\yyyyy$ and a a density matrix $\rrh$ in the state space of 
$\yyyy$, $\rrh$ is called {\dff $\pttt$-product} whenever it can be  
represented as a tensor product
\[
\rrh
=
\rrh_{\ptt_1}
\otimes\cdots\otimes
\rrh_{\ptt_M}
\] 
\noindent and {\dff $\pttt$-separable} if it is a convex 
combination of $\pttt$-product states:
\begin{equation}\label{e04}
\rrh
=
\sum
p_\alpha
\rrh_{\ptt_1}^\alpha
\otimes\ldots\otimes
\rrh_{\ptt_\lth}^\alpha
\end{equation}
\noindent In other words, \eqref{e04} means that we can prepare 
$\rrh$ as an ensemble of mixed states located at sites 
$\ptt_1,\ldots,\ptt_M$. 

\medskip 

Given a state $\rrh$, we  may now ask for each partition $\pttt$ of 
the set $\yyyyy$ of subsystems of $\yyyy$ if $\rrh$ is 
$\pttt$-separable or not. As a result we  obtain the set 
$\pttr(\rrh)$ of partitions of $\yyyyy$ with respect to which 
$\rrh$ is separable \eqref{e04}: 

\begin{equation}\label{e05}
\pttt\in\pttr(\rrh)
\quad\Leftrightarrow\quad
\rrh
\quad
\mbox{is $\pttt$-separable}
\end{equation}

\noindent For a given state $\rrh$ the set $\pttr(\rrh)$ has the 
following property \cite{durcirac}: 

\begin{equation}\label{e05n}
\forall \:\rrh
\quad
\left\lbrace
\begin{array}{l}
\pttt\in\pttr(\rrh)
\cr
\pttt'\preceq\pttt
\end{array}
\right.
\quad\Leftrightarrow\quad
\pttt'\in\pttr(\rrh)
\end{equation}

\paragraph{Corollary.} To 
specify $\pttr(\rrh)$ we need to provide only maximal (with respect 
to the order \eqref{e01}, that is, finest) partitions $\pttt$ such 
that $\rrh$ is $\pttt$-separable. This provides us a natural 
geometrical picture which will be coined partition polytopes. For 
the sake of self-consistency, I first recall the necessary 
definitions. 

\subsection{Simplices and polytopes}\label{sssympol}

Suppose we have a real affine space $\afs$ of sufficiently high 
dimension, and there are $M+1$ independent points there. The  
convex hull of these points is called {\dff simplex} of dimension 
$M$. For example: 

\setlength{\unitlength}{1mm}
\begin{itemize}
\item[] A cut 
\begin{picture}(20,4)
\put(5,0){\circle*{2}}
\put(15,0){\circle*{2}}
\put(6,0){\line(1,0){8}}
\end{picture}
is a 1-simplex.

\item[] A triangle
\begin{picture}(20,7)
\put(5,0){\circle*{2}}
\put(15,0){\circle*{2}}
\put(10,5){\circle*{2}}
\put(6,0){\line(1,0){8}}
\put(6,1){\line(1,1){4}}
\put(14,1){\line(-1,1){4}}
\end{picture}
is a 2-simplex.

\item[] A tetrahedron
\begin{picture}(20,11)
\put(5,0){\circle*{2}}
\put(15,0){\circle*{2}}
\put(10,5){\circle*{2}}
\put(10,10){\circle*{2}}
\put(6,0){\line(1,0){8}}
\put(6,1){\line(1,1){4}}
\put(5,1){\line(1,2){4}}
\put(14,1){\line(-1,1){4}}
\put(15,1){\line(-1,2){4}}
\put(10,6){\line(0,1){4}}
\end{picture}
is a 3-simplex.

\end{itemize}

\noindent and so on. The points we have connected are called {{\dff 
vertices} of the simplex. The convex hulls of $K+1$ vertices in a 
simplex are called its {\dff faces}. In particular, 0-faces are 
vertices, 1-faces  are called edges. 

\medskip 

A {\dff polytope} is a collection $\{\sss^i\}$ of simplices in 
the space $\afs$ having the following property: 

\begin{equation}\label{e02p}
\forall\,i,i'\qquad
\sss^i
\cap
\sss^{i'}
\:
=
\quad
\left[
\begin{array}{l}
\mbox{a face of both}
\quad
\sss^{i},\,
\sss^{i'}
\cr
\:\emptyset
\end{array}
\right.
\end{equation}

\noindent The {\dff vertices} of a  polytope is the set of all 
vertices of all its simplices, denote it by $\vvv$

\[
\vvv
=
\cup_i
\vvv(\sss^i)
\] 

\medskip 

\noindent For instance

\setlength{\unitlength}{1mm}
\begin{itemize}
\item[] 
\begin{picture}(20,4)
\put(5,0){\circle*{2}}
\put(15,0){\circle*{2}}
\put(6,0){\line(1,0){8}}
\end{picture},
\begin{picture}(20,11)
\put(5,0){\circle*{2}}
\put(15,0){\circle*{2}}
\put(5,10){\circle*{2}}
\put(15,10){\circle*{2}}
\put(6,0){\line(1,0){8}}
\put(6,10){\line(1,0){8}}
\put(5,1){\line(0,1){8}}
\put(15,1){\line(0,1){8}}
\put(6,1){\line(1,1){8}}
\end{picture},
\begin{picture}(30,11)
\put(5,0){\circle*{2}}
\put(5,10){\circle*{2}}
\put(15,10){\circle*{2}}
\put(25,0){\circle*{2}}
\put(25,10){\circle*{2}}
\put(6,10){\line(1,0){18}}
\put(5,1){\line(0,1){8}}
\put(25,1){\line(0,1){8}}
\put(6,1){\line(1,1){8}}
\put(24,1){\line(-1,1){8}}
\end{picture}
are polytopes
\vspace{3ex}

\item[]  while
\begin{picture}(30,11)
\put(5,0){\circle*{2}}
\put(15,0){\circle*{2}}
\put(15,10){\circle*{2}}
\put(6,0){\line(1,0){8}}
\put(6,1){\line(1,1){8}}
\put(15,1){\line(0,1){8}}
\put(5,5){\circle*{2}}
\put(25,5){\circle*{2}}
\put(6,5){\line(1,0){18}}
\end{picture},
\begin{picture}(20,11)
\put(5,0){\circle*{2}}
\put(15,0){\circle*{2}}
\put(5,10){\circle*{2}}
\put(15,10){\circle*{2}}
\put(6,0){\line(1,0){8}}
\put(6,1){\line(1,1){8}}
\put(5,1){\line(0,1){8}}
\put(15,1){\line(0,1){8}}
\put(14,1){\line(-1,1){8}}
\end{picture}
are not (provided $\dim\afs=2$)
\end{itemize}

Let $\sss$, $\sss'$ be two simplices in affine spaces $\afs, 
\afs'$. An affine mapping $\phi:\afs\to\afs'$ is called {\dff 
simplicial} whenever the following conditions hold: 

\begin{itemize}
\item Any vertex $v$ of $\sss$ is mapped onto a vertex of $\sss'$. 
This mapping is not necessarily injective, different vertices may 
be  mapped to one. 
\item `Simplicial mappings do not brake skeletons': if 
$\{v_1\cdt{}v_K\}$ is a face of $\sss$ then its image is a face as 
well (of perhaps smaller dimensionality). 
\end{itemize}

\noindent Both these conditions can be symbolically written down 
as: 

\begin{equation}\label{edefsm}
\phi(\sss) \subseteq \sss'
\end{equation}

\section{Topological expression of physical properties}

Now we have  in our disposal all the necessary mathematics to 
express certain physical properties of a  state $\rrh$ in a 
topological fashion. First I introduce the appropriate objects 
associated with states, namely, the \underline{separability polytopes} 
and then provide some examples, both trivial and non-trivial. 

\subsection{Separability polytopes}

Let us see how the introduced geometrical objects can 
represent relations between different partitions. With any 
partition $\pttt_\alpha$ of the  set $\yyyyy$ we associate a  
simplex whose vertices are in 1--1 correspondence with the 
elements of the partition $\pttt$ (= disjoint subsets of $\yyyyy$). 

\[
\sss_\alpha = 
\sss(\pttt_\alpha)
\] 

\medskip 

\noindent The separability polytope of a state $\rrh$ of $\yyyy$ is 
built as follows: 

\begin{itemize}
\item[(i)]
Take all maximal partitions $\{\pttt^1,\ldots,\pttt^K\}$ with 
respect to which $\rrh$ is separable. 

\item[(ii)]
Consider the set $\vvv$ whose elements are all elements of the 
partitions $\{\pttt^1,\ldots,\pttt^K\}$ (that is,  subsets of 
$\yyyyy$).  Formally it is just:

\[
\vvv
=
\{\pttt^1\cup\cdots\cup\pttt^K\}
\] 

\noindent and denote by $D$ the cardinality of the set $\vvv$. 

\item[(iii)]
Take an affine space $\afs$ of sufficiently high dimension and 
place there $D$ independent points corresponding to the elements of 
$\vvv$. They will be the vertices of the future separability 
polytope.  

\item[(iv)]
For each partition $\pttt^i$ form the convex hull of appropriate 
vertices. 
\end{itemize}

The result we call the {\dff separability polytope} of the state 
$\rrh$, denote it $\kkk(\rrh)$: 

\begin{equation}\label{edefsp}
\kkk
=
\cup
\{
\sss(\pttt^i)
\:\mid\:
\pttt^i\in\pttr(\rrh)
\}
\end{equation}

\bigskip 

Yet separability polytopes remain an abstract geometrical object 
for which physical interpretation is to be provided. A quantum 
state in general does not contain complete information about an 
ensemble which represents this state (again, this is not the case 
for classical systems where the correspondence state--ensemble is 
unique). What tells us the fact that a state $\rrh$ is separable 
with respect to a partition $\pttt=\{\ptt_1\cdt\ptt_M\}$? This 
means that $\rrh$ can be represented as a  mixture of pure states 
of the form:

\[
\rrh
=
\sum
p^\alpha
\ketbra{\Psi^\alpha_{\ptt_1}}{\Psi^\alpha_{\ptt_1}}
\otimes\cdots\otimes
\ketbra{\Psi^\alpha_{\ptt_M}}{\Psi^\alpha_{\ptt_M}}
\] 

\noindent If the  partition $\pttt$ is maximal, that means that 
each $\ket{\Psi^\alpha_{\ptt_i}}$ is entangled. So, the  simplex 
associated with a  maximal partition shows us the 
`localisation of entanglement' for ensembles representing the 
state $\rrh$. If there are several maximal partitions, each of 
these partitions is associated with a class of ensembles which can 
represent $\rrh$. Overlapping simplices indicate how similar are 
particular localizations of entanglement for different locally 
inequivalent ensembles representing the same state $\rrh$. Now 
consider some special cases, beginning with trivial ones where 
separability polytopes degenerate.

\subsection{Trivial cases}

Begin with pure states. If $\rrh$ is a pure state, then its 
separability polytope $\kkk(\rrh)$ is a simplex. Indeed, if $\rrh$ 
is pure, then it is a projector on a pure state 
$\ket{\Psi}\in\hhhh$. Let $\pttt=\{\ptt_1,\ldots,\ptt_M\}$ is such 
a partition of $\yyyyy$ that $\ket{\Psi}$ decomposes into a product

\begin{equation}\label{e07i}
\ket{\Psi}
=
\ket{\Psi_{\ptt_1}}
\otimes\cdots\otimes
\ket{\Psi_{\ptt_M}}
\end{equation}

\noindent such that each $\ket{\Psi_{\ptt_i}}$ is entangled. Then 
suppose that there is another partition 
$\pttt'=\{\ptt_1',\ldots,\ptt_M'\}$ incomparable with $\pttt$ in 
the sense \eqref{e01}. Then it turns out that $\ket{{\Psi}}$ has in 
addition to the decomposition \eqref{e07i} another one 

\begin{equation}\label{e07j}
\ket{\Psi}
=
\ket{\Psi_{\ptt_1'}}
\otimes\cdots\otimes
\ket{\Psi_{\ptt_M'}}
\end{equation}

\noindent Then tracing out $\ket{\Psi}$ expressed as \eqref{e07i} 
with respect to a subsystem $\ptt_i$ which is not in $\pttt'$ (it 
must exist as $\pttt$, $\pttt'$ are incomparable) we get a pure 
state. On the other hand, doing the same operation with $\ket{\Psi}$ 
expressed as \eqref{e07j} we get a mixed state. Therefore the 
conjectured $\pttt'$ does not exist, and we have the only 
partition $\pttt$ for the state $\rrh$. 

The separability simplex for a pure state is obtained as follows. 
We take the  simplex whose vertices are the subsystems of $\yyyy$ 
and collapse the vertices which belong to the same  element of the  
partition $\pttt$. In particular, when $\rrh$ is fully entangled, 
the polytope collapses to a single point. This shows that the 
topological classification we provide is not complete with respect 
to LOCC (local operations + classical communications) applied to a  
single copy of the system in question. It follows, for instance, 
from the existence of LOCC-inequivalent fully entangled tripartite 
mixed states \cite{AJDV00,DVC00,VDD01d}. 

However, for the special case of pure states of multi-qubit systems 
a more subtle classification exists based on the 
Carteret-Higuchi-Sudbery pure states decomposition \cite{chs00}. 
This classification starts from the same simplex of subsystems but 
is more `sparing' and does not collapse to the extent proposed  
here. For further details the reader is referred to \cite{RZ02}. 

The next trivial case is that of bipartite systems. It is 
degenerate for the proposed construction as the variety of 
entanglement patterns is quite poor, that is, a bipartite state 
can be either separable or entangled and there are no intermediate 
cases. 

\subsection{Three qubits.} 

Three-qubit systems provide us first non-trivial examples of 
essentially different entanglement patterns for mixed states. 
Although the number of subsystem is still small to provide 
`interesting' polytopes. Referring to the classification 
\cite{durcirac} of mixed 3-qubit states we have 5 different classes 
of states displayed in table \ref{tab1}

\begin{table}[h!]\label{tab1}
\setlength{\unitlength}{1mm}
\begin{tabular}{|c|p{0.75\textwidth}|}
\hline
\begin{picture}(20,10)
\put(10,0){\circle*{2}}
\end{picture}
&Fully entangled states for which the separability polytope 
shrinks to a single point. An example of such state is, say GHZ 
state \cite{ghz89}. 
\cr\hline
\begin{picture}(20,5)
\put(5,0){\circle*{2}}
\put(15,0){\circle*{2}}
\put(6,0){\line(1,0){8}}
\end{picture}
&
1-qubit biseparable states which can be separated as 
$\pttt=\{A,BC\}$ or $\pttt=\{B,AC\}$ or $\pttt=\{C,AB\}$. For them 
the separability polytopes are 1-simplices, that is, two points 
connected by an edge (see  section \ref{sssympol}). An example of 
such state may be a product state of a pure 1-qubit state and fully 
entangled state of the remaining two qubits. 
\cr\hline
\begin{picture}(20,5)
\put(5,0){\circle*{2}}
\put(15,0){\circle*{2}}
\put(6,0){\line(1,0){8}}
\put(5,-5){\circle*{2}}
\put(15,-5){\circle*{2}}
\put(6,-5){\line(1,0){8}}
\end{picture}
&
2-qubit biseparable states which can be separated as 
$\pttt=\bigl\{\{A,BC\},\{B,AC\}\bigr\}$ and similar ones where 
$B,C$ are not separated out, respectively. The explicit examples 
of such states can be found in \cite{durcirac}. For them the 
separability polytopes are disjoint unions of two 1-simplices. 
\cr\hline
\begin{picture}(20,5)
\put(5,0){\circle*{2}}
\put(15,0){\circle*{2}}
\put(6,0){\line(1,0){8}}
\put(5,-5){\circle*{2}}
\put(15,-5){\circle*{2}}
\put(6,-5){\line(1,0){8}}
\put(5,-10){\circle*{2}}
\put(15,-10){\circle*{2}}
\put(6,-10){\line(1,0){8}}
\end{picture}
&
An interesting and slightly counterintuitive example 
\cite{bvms99} is that of 3-biseparable states which are separable 
with respect to any bipartite split. For them the separability 
polytopes are disjoint unions of three 1-simplices.
\cr\hline
\begin{picture}(20,7)(0,5)
\put(5,0){\circle*{2}}
\put(15,0){\circle*{2}}
\put(10,5){\circle*{2}}
\put(6,0){\line(1,0){8}}
\put(6,1){\line(1,1){4}}
\put(14,1){\line(-1,1){4}}
\end{picture}
&
Finally, separable states provide us one more trivial 
example - for them the separability polytopes are 2-simplices (that 
is, triangles). 
\cr\hline
\end{tabular}
\caption{An exhaustive classification of entanglement 
patterns for the mixed states of a 3-qubit system.}
\end{table}

\paragraph{Remark.} In order to avoid confusion I emphasise that 
the lines drawn above show that the appropriate parties are 
\underline{separable} rather than entangled since usually, when 
entanglement pictures are drawn, the lines denote entangling 
interactions. 

\subsection{Physical meaning of separability 
polytopes}\label{ssphysmean}

Any mixed state $\rrh$ of an ensemble can be represented as a  
mixture of pure states. For classical systems this representation 
is unique. In contrast, any mixed state of a quantum system can be 
represented as an ensemble of pure states in infinitely many 
inequivalent ways. When the system in question is composite, 
multipartite, for any given particular ensemble $\ensm$ 
representing the state $\rrh$ we may ask: `is any of pure states 
forming $\ensm$ separable or not with respect to this or that 
partition of the set of subsystems $\yyyyy$ of the system $\yyyy$'? 
Or, in other words, we query if we  can represent the 
density operator as a mixture of appropriate product 
states. Separability polytopes are exactly the objects which capture 
this property. They show how many inequivalent (from the point of 
view of localisation) ensembles can form a given state $\rrh$. 

\paragraph{`Relatively local' transformations} What can 
separability polytopes also serve  for,  is to broaden the group of 
unitary transformations with respect to which the entanglement 
picture of a multipartite state remains unchanged. We can derive it 
using the  fact that separability polytopes are not given ad 
hoc arbitrary ones. Their maximal simplices are associated with 
partitions. Recall that the collection of all possible partitions 
of a set $\yyyyy$ is partially ordered by the relation \eqref{e01}. 
Moreover, the set of partitions has the structure of a lattice 
\cite{lt}, that is, for any two partitions there exist their least 
upper bound and the greatest lower bound with respect to relation 
\eqref{e01}. This means, in turn, that we have  a well defined 
operations of {\dff join} $\lor$ and  {\dff meet} $\land$ on 
simplices: 

\[
\sss_1\lor\sss_2\;=\;\sup\{\sss_1,\sss_2\}
\]
\[
\sss_1\land\sss_2\;=\;\inf\{\sss_1,\sss_2\}
\]

\noindent In the sequel we shall be interested only in $\lor$ 
operation. Let $\sss=\sup\{\sss_i\}$ and consider the class 
(a group, in fact) $\relloc_\rrh$ of unitary 
transformation which are local with respect to the partition 
$\sss$ which is the refinement of \underline{all} maximal 
simplices (=partitions) of the separability polytope $\kkk(\rrh)$. 
Then any transformation from this group applied to $\rrh$ keeps its 
separability polytope $\kkk(\rrh)$ unchanged, however $\relloc_\rrh$ 
is in general greater than the group of `genuinely' local 
transformations. 

\section{Implementations of quantum gates from topological 
perspective}

In this section I consider a register $\yyyy$ of $\lth$, say, qubits 
(the number of individual degrees of freedom is not important) and 
stepwise implementation of an arbitrary unitary transformation on 
it. By `stepwise' I mean that at each time step only a 2-qubit 
transformation is carried out. Proceeding this way,  one can 
implement, having a very limited scope of resources in disposal, an 
arbitrary unitary transformation on the whole  system $\yyyy$ up to 
any given accuracy \cite{gates95}.

Now suppose we have a  compound system $\yyyy$. In the sequel call 
it \underline{register}, and call its subsystems 
\underline{qubits}. Let us see what happens with the separability 
polytope of the state of the register in course of the stepwise 
execution of this simulation. We start with a certain initial (in 
general, mixed) state $\rrh_0$ with the separability polytope 
$\kkk(\rrh_0)$. As it was claimed in section \ref{ssphysmean}, 
$\kkk(\rrh_0)$ represents the equivalence classes of ensembles 
which can realise the state $\rrh_0$. At first step, when we pass 
from $\rrh_0$ to $\rrh_1=\uu_0\rrh_0\uu_0^\dag$ we may view it as 
follows: pick an ensemble $\ensm$ of pure states associated with a 
particular maximal simplex of $\kkk(\rrh_0)$ and apply $\uu_0$ to 
each pure state forming $\ensm$. 

For general $\uu_0$ nothing certain can be said, but recall that we 
consider a special kind of transformations, namely, entangling at 
most two qubits. Let $\ensm$ be an ensemble representing $\rrh$, 
associate with it a partition $\pttt$ of $\yyyy$. The action of 
$\uu_0$ may be either (i) entangling two qubits within one element 
of the covering associated with the a partition or (ii) `breaking 
the border', that is, entangling two qubits from different elements 
of the partition. Both these options have direct topological 
interpretation. 

In case (i) the appropriate partition remains intact, in case (ii) 
two elements of the partition merge. In terms of simplices option 
(ii) means that an edge of appropriate simplex collapses to a 
vertex. In both cases  this transition is a simplicial mapping 
(see section \ref{sssympol}). To see the real action of $\uu_0$ we 
must apply it to all non-equivalent ensembles (= maximal simplices 
of $\kkk(\rrh_0)$). As a result, the whole complex undergoes a 
simplicial contractin---some  edges of $\kkk(\rrh_0)$ may collapse 
to vertices. 

Since the vertices of $\kkk(\rrh_0)$ are subsets of the set 
$\yyyyy$ of subsystems of $\yyyy$, it may happen that the same 
subset will be obtained as different junctions. For example, 
suppose we have the following edges (= pairs of disjoint subsets)

\[
e_1 = \{\{1,2\},\{3\}\}\,, \qquad
e_2 = \{\{1\},\{2,3\}\}
\]

\noindent and we see that they have no common vertices. When they 
both collapse, we get 

\[
\uu_0(e_1) = \{1,2,3\}\,, \qquad
\uu_0(e_2) = \{1,2,3\}
\]

\noindent which means  that in the resulting complex the two 
vertices $\uu_0(e_1), \uu_0(e_2)$ should be identified. And again,  
this operation never `breaks edges' that is why it is a  simplicial 
mapping. 

So, any two-qubit unitary transform on a composite system is 
associated with a simplicial mapping of appropriate separability 
polytopes. But, as it was mentioned above, \underline{any} unitary 
transformation can be implemented as a sequence (in general 
infinite) of 2-qubit transformations. 

Now recall that we are dealing with finite-dimensional systems for 
which any set of states sharing the same entanglement pattern is 
of non-zero measure. That means  that we can arrange the 
implementation in such a way that after a finite number of steps 
the separability polytope of the  state will not change any longer. 
As a result, taking into account that compositions of simplicial 
mappings are always simplicial, we can conclude that 

\begin{quote}
Any unitary transformation of any given state of a 
finite-dimensional multipartite system can be associated with a 
simplicial mapping of their separability polytopes. 
\end{quote}

\paragraph{Remark.} Note that if we consider a unitary 
transformation \underline{per se}, not implemented as a sequence of 
at most 2-qubit operations, it has in general nothing to do with 
simplicial mappings. Therefore a particular simplicial mapping is 
associated with a \underline{realisation} of the transformation 
rather than with the transformation itself. 

\section*{Concluding remarks}

I have introduced  an essentially non-numerical characterisation of 
mixed states of multipartite quantum systems. With each state 
$\rrh$ a polytope---purely topological object which can be realised 
as a polyhedron in a Euclidean space - is associated. These 
polytopes capture the localisation properties of pure states 
forming ensembles which can represent the state $\rrh$. 
(by localisation of a pure state I mean with respect to which 
groups of subsystems it is entangled). 

It turns out that the presented characterisation is related with 
the simulation of arbitrary unitary operations by 1- and 2-qubit 
gates. Namely, to each particular implementation of such kind a 
simplicial mapping---a well known construction in algebraic 
topology---can be put into correspondence. 

\paragraph{Acknowledgments.} I am grateful to the members of 
Quantum Information Group, ISI  for permanent attention to the work 
and many helpful comments and discussions. Valuable comments from 
Demosthenes Ellinas, Markus Grassl and Philippe Jorrand are 
appreciated. I would like to thank the referee of Journal of 
Modern Optics for essential remarks and suggestions. The work was 
carried out under the auspices of the EU grant Q-ACTA `Quantum 
computation: novel algorithms and their many-body implementations' 
(IST-1999-10596) and Russian Fundamental Research Foundation 
(RFFI). 

\bigskip

\begin{flushleft}

\end{flushleft}

\end{document}